%% file: main.tex
\documentclass[conference]{IEEEtran}
\input{./tex_content/UsePackages.tex}
\input{./tex_content/macros.tex}

\usepackage[colorinlistoftodos]{todonotes}

\newcommand{\sysname}{\textit{MaxRay}}
\title{MaxRay: A Raytracing-based Integrated Sensing and Communication Framework}

\author{
\IEEEauthorblockN{M.~Arnold\IEEEauthorrefmark{1}, M.~Bauhofer\IEEEauthorrefmark{2}, S.~Mandelli\IEEEauthorrefmark{1}, M.~Henninger\IEEEauthorrefmark{1}, F.~Schaich\IEEEauthorrefmark{1}, T.~Wild\IEEEauthorrefmark{1}, S.~ten Brink\IEEEauthorrefmark{2}}
\IEEEauthorblockA{
\IEEEauthorrefmark{1}Nokia Bell Labs, Magirusstr. 8, 70469 Stuttgart, Germany
\\\{maximilian.arnold,  silvio.mandelli, marcus.henninger, frank.schaich, thorsten.wild\}@nokia-bell-labs.com
}
\IEEEauthorblockA{
\IEEEauthorrefmark{2}Institute of Telecommunications, Pfaffenwaldring 47, University of  Stuttgart, 70659 Stuttgart, Germany 
\\\{bauhofer, tenbrink\}@inue.uni-stuttgart.de
}
}

\begin{document}
% Input Acronyms to be used
\input{./tex_content/acronyms.tex}
% Make Title
\maketitle
\thispagestyle{empty}
\pagestyle{empty}

% Abstract
\begin{abstract}
\ac{ISAC} forms a symbiosis between the human need for communication and the need for increasing productivity, by extracting environmental information leveraging the communication network. 
As multiple sensory already create a perception of the environment, an investigation into the advantages of \ac{ISAC} compare to such modalities is required. Therefore, we introduce \textit{MaxRay}, an \ac{ISAC} framework allowing to simulate  communication, sensing, and additional sensory jointly.
Emphasizing the challenges for creating such sensing networks, we introduce the required propagation properties for sensing and how they are leveraged. To compare the performance of the different sensing techniques, we analyze four commonly used metrics used in different fields and evaluate their advantages and disadvantages for sensing. We depict that a metric based on prominence is  suitable to cover most algorithms.
Further we highlight the  requirement of clutter removal algorithms, using two standard clutter removal techniques to detect a target in a typical industrial scenario. 
In general a versatile framework, allowing to create automatically labeled datasets to investigate a large variety of tasks is demonstrated.
\end{abstract}

% Reset acronyms
\acresetall

% Main Matter
\input{./sections/introduction}

\input{./sections/framework}
\input{./sections/sensing}
\input{./sections/simulation}
\input{./sections/conclusion}

%\section*{Acknowledgements}
%We would like to thank Thomas Schlitter, Jürgen Otterbach, Stephan Saur and Artjom Grudnistky for their useful inputs and fruitful discussions.

\bibliography{biblio}
\bibliographystyle{IEEEtran}  

\end{document}

%% file: tex_content/UsePackages.tex
\usepackage{tikz}  
\usetikzlibrary{spy}
\usepackage{color, colortbl}
\usepackage{pgfplots}
\pgfplotsset{compat=newest}
\usepackage{multicol}
\usepackage{circuitikz}
\usepackage[nolist]{acronym}
\usepgfplotslibrary{groupplots}
\usetikzlibrary{shapes}
\usepackage{siunitx}  
\usepackage{graphicx}
\usepgfplotslibrary{patchplots}
\usepackage{amsmath,amssymb}
\usepackage{subfig}
 \usepackage{float}
\usepackage{cite}
\usepackage{amsmath}
\usetikzlibrary{angles, quotes,calc,patterns}
\usepgfplotslibrary{fillbetween}
\definecolor{mittelblau}{RGB}{0, 126, 198}
\definecolor{violettblau}{cmyk}{0.9, 0.6, 0, 0}
\definecolor{rot}{RGB}{238, 28 35}
\definecolor{apfelgruen}{RGB}{140, 198, 62}
\definecolor{gelb}{RGB}{1, 221, 0}
\definecolor{orange}{RGB}{244, 111, 33}
\definecolor{pink}{RGB}{237, 0, 140}
\definecolor{lila}{RGB}{128, 10, 145}
\definecolor{hellgrau}{RGB}{224, 224, 224}
\definecolor{mittelgrau}{RGB}{128, 128, 128}
\definecolor{dunkelgrau}{RGB}{80,80,80}
\definecolor{anthrazit}{RGB}{19, 31, 31}
\definecolor{darkgreen}{RGB}{0.125,0.5,0.169}

\usepackage{ifthen}
\usepackage{algorithm}
\usepackage[noend]{algpseudocode}
\usetikzlibrary{matrix,calc}
\usepackage{trfsigns}

%% file: tex_content/macros.tex
\renewcommand{\vec}[1]{\mathbf{#1}}

\newcommand{\av}{\vec{a}}
\newcommand{\bv}{\vec{b}}

\newcommand{\hv}{\vec{h}}

\newcommand{\pv}{\vec{p}}

\newcommand{\rv}{\vec{r}}

\newcommand{\tv}{\vec{t}}

\newcommand{\Hm}{\vec{H}}

\newcommand{\Pm}{\vec{P}}

\newcommand{\Rm}{\vec{R}}

\newcommand{\Xm}{\vec{X}}
\newcommand{\Ym}{\vec{Y}}

%% file: tex_content/acronyms.tex
\begin{multicols}{2}
\begin{acronym}[WSSUS]

    \acro{3GPP}{3rd Generation Partnership Project}
    \acro{5G}{fifth-generation}
    \acro{ADC}{analog to digital converter}
    \acro{AFE}{analog front end}
    \acro{AGC}{automatic gain control}
    \acro{AGV}{automated guided vehicle}
    \acro{AMP}{approximate message passing}
    \acro{API}{Application Programming Interface}
    \acro{AWGN}{additive white Gaussian noise}

    \acro{BER}{bit error rate}
    \acro{BB}{baseband}
    \acro{bpcu}{bits per channel use}
    \acro{BP}{belief propagation}
    \acro{BPSK}{binary phase shift keying}
    \acro{BS}{base station}

    \acro{CB}{codebook}
    \acro{CDF}{cumulative distribution function}
    \acro{CFO}{carrier frequency offset}
    \acro{CoSaMP}{compressive sampling matching pursuit}
    \acro{CP}{cyclic prefix}
    \acro{CS}{compressive sensing} 
    \acro{CSI}{channel state information}
    \acro{CNN}{convolutional neural network}

    \acro{DA}{domain adaptation}
    \acro{DAC}{digital-analog-converter}
    \acro{DC}{direct current}
    \acro{DE}{distance error}
    \acro{DeepL}{deep-learning}
    \acro{DoF}{degree-of-freedom}
    \acro{DFT}{discrete Fourier transformation}
    \acro{DL}{deep learning}
    \acro{DS}{delay spread}
    \acro{DSP}{digital signal processing}

    \acro{ECC}{error-correcting code}
    \acro{ENoB}{effective number of bits}
    \acro{ERP}{effective radiated power}
    \acro{EVM}{error vector magnitude}
    \acro{EVD}{eigenvector decomposition}
    \acro{FB}{feedback}
    \acro{FC}{fully connected}
    \acro{FDD}{frequency division duplexing}
    \acro{FDM}{frequency division multiplexing}
    \acro{FIR}{finite impulse response}
    \acro{FFT}{fast fourier transform}
    \acro{FT}{fine tuning}
    \acro{FPGA}{field programmable gate array}
    \acro{GAN}{Generative adversarial network}
    \acro{GPIO}{general-purpose input/output}
    \acro{GPS}{global positioning system}
    \acro{GPSDO}{GPS disciplined oscillator}
    \acro{GPU}{graphical processing unit}
    \acro{HDF}{Hierarchical Data Format}
    \acro{HDD}{hard decision decoding}
    \acro{IC}{integrated circuit}
    \acro{ICI}{inter-carrier-interference}
    \acro{ISAC}{Integrated Sensing And Communication}
    \acro{I2C}{Inter-Integrated Circuit}
    \acro{ICSP}{in-circuit serial programming}
    \acro{IF}{intermediate frequency}
    \acro{i.i.d.}{independent and identically distributed}
    \acro{IIR}{infinite impulse response}
    \acro{IMU}{inertial measurement unit}
    \acro{IoT}{Internet of Things}
    \acro{IPS}{indoor positioning system}
    \acro{IR}{infrared}
    \acro{JSDM}{Joint Spatial Division and Multiplexing}
    \acro{LIDAR}{Light Detection And Ranging}
    \acro{LLR}{log-likelihood ratio}
    \acro{LP}{leakage precoder}
    \acro{LMMSE}{Linear Minimum Mean Square Error}
    \acro{LO}{local oscillator}
    \acro{LoS}{line of sight}
    \acro{LiDaR}{Light Detection and Ranging}
    \acro{LS}{least squares}
    \acro{LSTM}{long-term short-term memory}
    \acro{LTE}{Long Term Evolution}
    \acro{LTI}{linear time invariant}
    \acro{LTV}{linear time variant}
  
    \acro{MAP}{maximum a posteriori}
    \acro{MDE}{mean distance error}
    \acro{MDA}{mean distance accuracy}
    \acro{MEMS}{Micro-Electro-Mechanical Systems}
    \acro{MIMO}{multiple input multiple output}
    \acro{MISO}{multiple input single output}
    \acro{ML}{maximum likelihood}
    \acro{MLD}{maximum likelihood decoding}
    \acro{mMIMO}{massive multiple input multiple output}
    \acro{MMSE}{minimum mean square error}
    \acro{M-MMSE}{multi-cell minimum mean square error}
    \acro{MR}{maximum ratio}
    \acro{MRC}{maximum ratio combining}
    \acro{MRP}{maximum ratio precoding}
    \acro{MRT}{maximum ratio transmission}
    \acro{MSE}{mean squared error}
    \acro{MQTT}{Message Queuing Telemetry Transport}
    \acro{MU}{multi-user}
    \acro{MUSIC}{Multiple Signal Classification}
    \acro{NF}{noise figure}
    \acro{NN}{Neural Network}
    \acro{NNI}{Neural Network Intelligence}
    \acro{NLoS}{non-line of sight}
    \acro{NND}{neural network decoding}
    \acro{NTP}{Network Time Protocol}
    \acro{NMSE}{normalized mean squared error}
    \acro{NU}{not-used}
    \acro{OFDM}{orthogonal frequency division multiplex}
    \acro{OMP}{orthogonal matching pursuit}
    \acro{OPS}{outdoor positioning system}
    \acro{OT}{optimal transport}
    \acro{PB}{passband}
    \acro{PCB}{printed circuit board}
    \acro{PDR}{pedestrian dead reckoning}
    \acro{PDF}{probability density function}
    \acro{PDP}{power-delay-profile}
    \acro{PLL}{phase-locked-loop}
    \acro{PO}{phase-only}
    \acro{PPS}{pulse per second}
    \acro{QPSK}{quadrature phase shift keying}
    \acro{QuaDRIGa}{Quasi Deterministic Radio Channel Generator}

    \acro{RADAR}{Radio Detection And Ranging}
    \acro{ReLU}{rectified linear unit}
    \acro{RF}{radio frequency}
    \acro{RMS-DS}{Root Mean Square - Delay Spread}
    \acro{RNN}{recurrent neuronal network}
    \acro{RSSI}{received signal strength indicator}
    \acro{R-ZF}{regularized zero-forcing}
    \acro{SDD}{soft decision decoding}
    \acro{SDR}{software defined radio}
    \acro{SE}{spectral efficiency}
    \acro{SFO}{sampling frequency offset}
    \acro{STO}{sampling time offset}
    \acro{SLAM}{Simultaneous Localization and Mapping}
    \acro{SGD}{stochastic gradient descent}
    \acro{SISO}{single input single output}
    \acro{SINR}{signal-to-interference-and-noise-ratio}
    \acro{SIR}{signal-to-interference-ratio}
    \acro{SLNR}{signal-to-leakage-and-noise ratio}
    \acro{SNR}{signal-to-noise-ratio}
    \acro{SP}{subspace}
    \acro{SQR}{signal-to-quantization-noise-ratio}
    \acro{SQNR}{signal-to-quantization-noise-ratio}
    \acro{SVD}{singular value decomposition}
    \acro{SU}{single-user}
    \acro{TDD}{time division duplexing}
    \acro{TRIPS}{time-reversal IPS}
    \acro{UE}{user equipment}
    \acro{UL}{uplink}
    \acro{ULA}{uniform line array}
    \acro{URLLC}{ultra-reliable low-latency communication}
    \acro{US}{uncorrelated scattering}
    \acro{USRP}{universal software radio peripheral}
    \acro{UWB}{ultra-wideband}
    \acro{WiFi}{Wireless Fidelity}
    \acro{WSS}{wide sense stationary}
    \acro{WSSUS}{wide sense stationary uncorrelated scattering}

    \acro{ZF}{zero forcing}
\end{acronym}
\end{multicols}

%% file: sections/introduction.tex
\section{Introduction}\label{Sec:Introduction}
%% Sensing was always there and was not called like it
\ac{ISAC} is widely seen as the natural evolution of of communications-only networks. In the next generation of wireless network standards, \ac{ISAC} will enable the already existing communication links to extract environment information~\cite{TWild2021,Lima2021}. The synergy between communications and sensing could be then leveraged~\cite{viswanathan2020communications}. Although sensing is seen as a new topic in the world of communication, we want to emphasize that both worlds are learning from each other,  e.g., \ac{RADAR} applications use communication principles like \ac{MIMO} and \ac{OFDM} \cite{Braun2019,Leen2012}. On the other hand, proof of concepts where communication networks can locate their users have already been published~\cite{saur20205gcar}.
However, not only active users can be located, but wireless link can also be used to detect passively the activity of humans \cite{Jian2020}, their gestures \cite{Wenfeng2015}, or their position \cite{Arnold2019}.

%% A redesign and rethinking of both worlds needs to be happening
Although seeing the two worlds merging towards a joint system, a complete redesign of them to fully unveil their potential is still to be done. 
Among the open points, the channel models typically considered for communications are statistical~\cite{3GPP}. The main shortcoming of these models is that they not consider the interaction among the environment and the signals propagating through it in a deterministic way, thus not allowing to perform any meaningful investigation about sensing.
Therefore, raytracing can be considered allowing to model the deterministic wireless network signal's propagation in the environment. Moreover, raytracing offers the capabilities to reproduce the response of additional sensors, like \ac{LIDAR} and cameras. This allows to investigate not only \ac{ISAC}, but also sensor fusion techniques.

\begin{figure}[t]
    \centering
        \includegraphics{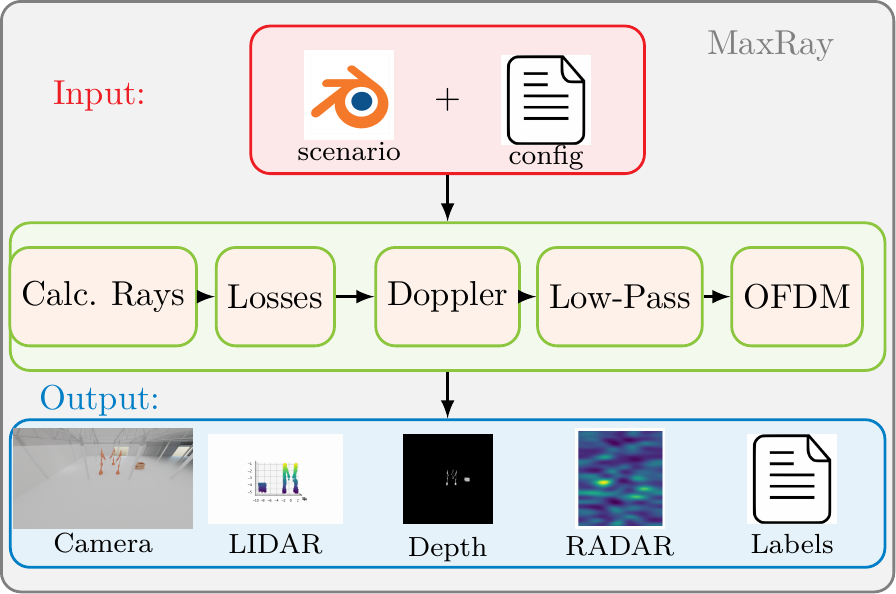}
    \caption{Blockdiagram of the MaxRay Framework}
    \label{fig:framework}    \vspace{-0.5cm}
\end{figure}

%% Out proposal MaxRay
Therefore, we introduce \sysname, a versatile tool to simulate realistic scenarios, leveraging ray-tracing to get \ac{ISAC} channel responses. Moreover, MaxRay allows to deploy and generate the acquisition of multiple environmental sensors, for the moment cameras and \ac{LIDAR}.
The proposed \sysname is used in an indoor factory scenario to generate an extensive dataset available to the community\footnote{Link available after review} for evaluating \ac{ISAC} and sensor fusion experiments. 
In this paper, we leverage the generated data to address a well-known problem in \ac{RADAR} literature, i.e. clutter removal. Indeed, a \ac{RADAR} acquisition comes including all the environment information that may disturb the sensing tasks. For instance, in our considered \ac{AGV} tracking scenario, we detect also factory walls and static equipment, that are defined as clutter, in addition to the \ac{AGV}. One would like to remove the clutter to have clean acquisitions, improving the sensing-based \ac{AGV} tracking performance.
Therefore, we systematically evaluate two standard clutter removal techniques with four different metrics used in computer vision, \ac{RADAR} and/or communication systems. The sensitivity of these metrics with respect to the \ac{AGV} tracking precision is discussed, providing suggestions on which are the metrics to be evaluated for network sensing applications.

Summarizing our contributions
\begin{itemize}
    \item we discuss the viability of raytracing to investigating \ac{ISAC},
    \item we investigate clutter removal performance in an indoor factory scenario and \ac{ISAC} parametrization, defining and assessing different metrics to measure the performance,
    \item we provide the labeled dataset to reproduce our results. The dataset can be used for other \ac{ISAC} and/or sensor fusion studies.
\end{itemize}

%% file: sections/framework.tex
\section{MaxRay: A versatile \ac{ISAC} Framework} \label{sec:framework}
Fig.~\ref{fig:framework} depicts the workflow of \sysname, where the input is a dynamic scenario encapsulated as a Blender\cite{Blender} file and  configuration file. This configuration file contains the communication settings, e.g. carrier frequency and number of subcarriers. The power of using Blender is modeling realistic environments, where fine-granular movement of parts and complex scenarios can be simulated.
Exploiting the Python \ac{API} \cite{Blender} the exact information about the environment can be leveraged by a ray tracer, getting the communication channel impulse response. Thus, we give a rough overview of the pipeline before explaining the parts in detail.

First, all possible paths considering different propagation properties (reflection, scattering, diffraction, blockage) are calculated using this \ac{API} and their interactions based on the incident angle, outgoing angle, materials, and speed of the surface are denoted. This computationall< complex process is done within the Blender environment and then passed through the following routine within \textit{MaxRay}:
Leveraging those parameters, all experienced losses per path (reflection loss, penetration loss,..) are calculated. Then, a Doppler component (phase shift) per path is added, corresponding to the movement of each interacted frame. Further the \ac{OFDM} frame structure is superimposed by band-limiting the signal and applying the Fourier-transform.
On this \ac{OFDM} frame the corresponding sensing/\ac{RADAR} images can be calculated. 

In general the outputs of \textit{MaxRay} can be a rendered Camera, a depth and/or a \ac{LIDAR} image containing all back-scattering points and their distance. Further, the bounding boxes for the camera, the \ac{LIDAR} object identification per point, the depth bounding boxes and the original position and dimensions of the objects in the environment, are created automatically.  We are emphasizing that a full pipeline for different \ac{DL} or classical techniques is created and the datasets will be available to reproduce our results.

\subsection{Geometrical RayTracing}
\begin{figure}[H]
    \centering
        \includegraphics{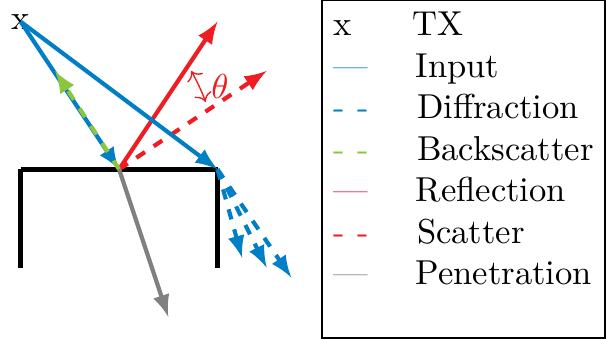}
    \caption{Different interactions with objects}
    \label{fig:Interaction-with-Objects}     %\vspace{-0.4cm}
\end{figure}

\begin{algorithm}[t]
\caption{Geometrical ray-casting}\label{alg:cap}
\begin{algorithmic}
\Require probing vectors $\Pm$, max interactions $\ell$, RX pos $\rv$, TX pos $\tv$, Current number interaction $i = 0$ 
\State \empty
\Procedure{trace-path}{$\Pm,\tv,\rv,i,\ell$}
    \If{$i \leq \ell$} \Comment max interaction not reached
        \For{$\pv$ in $\Pm$}
            \If{check-if-rx-hit($\tv,\rv,\pv$)}
                \State store-trace();
                \State break;
            \EndIf
            \State hit,\textbf{point},obj $\gets$ ray($\tv$,$\pv$) 
            \If{hit}
                \State $i \gets i +1$;
                \State $\tv \gets \textbf{point}$;
                \State $\hat{\Pm} \gets$ calc-new-probes($\pv$,obj);
                \State trace-path($\hat{\Pm},\tv,\rv,i,\ell$); \Comment trace recursively
            \EndIf
        \EndFor 
    \EndIf
\EndProcedure
\end{algorithmic}    
\end{algorithm}
The pseudo-code of the geometrical ray-casting is given in Alg.~\ref{alg:cap}. The input is a set of probing vectors  $\Pm$ seen from the transmitter point into the environment, compare the blue solid lines of Fig.~\ref{fig:Interaction-with-Objects} as an example.. Thus, $\Pm$ defines the spatial resolution and accuracy of the ray-tracing engine, as the initial number and resolution of rays are set by the user. Note in our case we set it to the full angle domain using a resolution of $1\textdegree$. Then, in a recursive fashion each ray is traced, allowing it to penetrate, diffract, back-scatter and reflect from the objects.
Thus, the function "ray()" calculates from the position $\pv$ into the direction $\tv$ the next interacting object, returning if and at which location ($\textbf{point}$) an interaction occurred.
From this location, the function "calc-new-probes()"  calculates using all propagation effects a new set of probing vectors, which will be explained in Fig.~\ref{fig:Interaction-with-Objects}. This path is stored and broken if at any point a bounding box at the receiver is hit, e.g. by calling the function "check-if-rx-hit". For this to be more efficient, a cube around the receiver of four times the wavelength $\lambda$ is considered. After computing each path, up to the $\ell$-th interaction, their corresponding delays and losses will be calculated.
Fig.~\ref{fig:Interaction-with-Objects} depicts for an input vector (blue)
\begin{itemize}
    \item the back-scattering component by inverting the direction of the input vector,
    \item the reflected vector by mirroring the input vector,
    \item the scattering vector created by a pre-defined angular spread of, e.g. 10 degrees and resolution of 1 degree a set of rays around he reflected ray,
    \item the penetration vector by using Snell's-law,
    \item the set of diffraction vectors if the input vector hits close to an edge.
\end{itemize}
The difference between the reflected and scattered probing vector is given by the angular spread $\theta$. From one ray as an input a manifold of new probing vectors emerge, allowing to model most common propagation effects. In the next subsection we elaborate on how to compute the losses for each of these interactions.

\subsection{Propagation Properties: Losses}
All traced paths undergo certain lossy interactions which are accounted for by
\begin{equation}
p_{\text{loss}} = \Gamma_{\text{path}}\Gamma_{\text{beam}} \Gamma_{\text{reflect}} \Gamma_{\text{scat}} \Gamma_{\text{rough}} \Gamma_{\text{diffract}} \Gamma_{\text{backscatter}},
\end{equation}
 multiplying all effects along the path. Hereby 
 \begin{itemize}
     \item  $\Gamma_{\text{path}}$ is the free-space path-loss,
     \item  $\Gamma_{\text{beam}}$ is the antenna pattern-loss,
     \item  $\Gamma_{\text{reflect}}$ is the reflection loss,
     \item  $\Gamma_{\text{scat}}$ is the additional scattering loss,
     \item  $\Gamma_{\text{rough}}$ is the roughness loss,
     \item  $\Gamma_{\text{diffract}}$ is the diffraction loss,
     \item  $\Gamma_{\text{backscatter}}$ is the back-scatter loss.
 \end{itemize}
 Emphasizing that a full explanation of these effects would go beyond the scope of this paper, we provide the equations we used, with some brief explanation.  Note that these models can easily be changed or improved and adjusted to different measurements.
The free-space path-loss
\begin{equation}
    \Gamma_{\text{path}} = \left (\frac{\lambda}{4\pi d}\right),
\end{equation}
 is commonly calculated by using the wavelength $\lambda$ and the distance travelled along the path between transmitter and receiver. The antenna pattern is used to calculate the beam loss for the transmitter and receiver pair and can be chosen from a perfect dipole/patch, a measured patch and a beamset for mmWave systems.
 The reflection loss 
\begin{equation}
    \Gamma_\text{reflect} = \frac{\cos{\phi}-\sqrt{\mu_r(f) \epsilon_r(f) - \sin^{2}{\phi}}}{\cos{\phi}+\sqrt{\mu_r(f) \epsilon_r(f) - \sin^{2}{\phi}}}
    \label{eg:reflect_loss}
\end{equation}
is due to the material change from air to a material with specific frequency-dependent permittivity $\epsilon_r$, permeability $\mu_r$, influenced by the input angle $\phi$. Note that if the material is a perfect reflector the reflection coefficient goes to one. The scattering loss \cite{Rappaport2019}
\begin{equation}
    \Gamma_\text{scatt} = \left(\frac{1+\cos{\theta}}{2}\right)^{\frac{\alpha_r}{2}},
\end{equation}
accounts for the effect that other rays emerge around the original reflected ray due to the roughness of the material. Hereby the $\alpha_r$ is a parameter to model the angular spread around the reflected ray\cite{Rappaport2019}.
Additionally to the scattering component a roughness loss \cite{Rappaport2019}
\begin{equation}
    \Gamma_\text{rough} = \exp\left[{-8\left(\frac{\pi\rho\cos{\phi}}{\lambda }\right)^2}\right] J_0\left[8\left(\frac{\pi\rho\cos{\phi}}{\lambda }\right)\right],
\end{equation}
is required, using the Bessel function of zero-th order $J_0$ and the standard deviation of the surface roughness $\rho$. This roughness loss is is to model the interaction of high frequency waves at surfaces where the roughness is comparable to the wavelength.
The diffraction loss
\begin{equation}
    \Gamma_{\text{diffract}}(\nu) = \frac{\sqrt{\left(1-C(\nu)-S(\nu)\right)^2 + \left(C(\nu) + S(\nu)\right)^2}}{2},
\end{equation}
is commonly modeled using the ITU-R P.526-14 standard \cite{ITU}.
Hereby we use the knife-edge diffraction by calculating the geometrical factor $\nu=\sqrt{\frac{2d}{\lambda}\alpha_1 \alpha_2}$, where the angles $\alpha_1,\alpha_2$ are between the top of the obstacle and one end as seen from the other end. Moreover, the Fresnel integral
\begin{equation}
    F(\nu) = C(\nu) + jS(\nu)
\end{equation}
 is approximated by using the cosine integral $C(\nu)$ and the sine integral $S(\nu)$. For more details we refer to \cite{ITU}.
The last effect considered is the back-scattering loss
\begin{equation}
    \Gamma_{\text{backscatter}} = p_\text{scat} \left(\frac{1+\cos{\phi}}{2}\right)^{\frac{\alpha_r}{2}},
\end{equation}
where we consider the input angle $\phi$ and the specific back scatter loss $p_\text{scat}$ as the main back-reflection component. One can derive that if the incoming angle is close to zero the effect of back-scattering is stronger than if the ray emerges from the side.

\subsection{Baseband channel representation}
Blender and its animation feature can be used to calculate the effective movement between animated frames, allowing to calculate the effective Doppler phase-shift
\begin{equation}
 \beta = 4\pi(N_\text{sub}+ \text{CP})f_s f_c \frac{v_s}{c_0}   
\end{equation}
per path. $N_\text{sub}$ is the number of subcarriers, \text{CP} the cyclic prefix length, $f_s$ the sampling rate, $f_c$ the carrier frequency, $c_0$ the speed of light in the vacuum and $v_s$ the relative speed in each path-element. This Doppler shift is applied per path at the receiver. 
As each transmission scheme has a certain bandwidth, the channel impulse response is now decimated. The wanted OFDM frames are created by zero-padding this band-limited channel impulse response and applying the Fourier transform.

The \ac{OFDM} frames, additional sensory and corresponding labels are  saved into a large \ac{HDF} file.
Although one would now expect measurements to verify the ray-tracing core we shift this investigation to a later point, emphasizing that we are not claiming to create a good match between measurements and simulation, but claiming that the underlying structure within behaves the same. 

%% file: sections/sensing.tex
\section{Sensing Basics} \label{sec:sens}
For understanding the full leverage of \ac{ISAC}, first the basics of sensing needs to be understood. Thus, we first explain the basic sensing concepts and show the challenges of such systems.
\subsection{OFDM \ac{RADAR}}
Considering that standard communication systems use \ac{OFDM}, we are going to exploit the concept of \ac{OFDM} \ac{RADAR} \cite{Braun2019} for sensing. Thus, knowing for each of the $N_\text{symb}$ transmitted symbols, the transmitted data  $\Xm \in \mathbb{C}^{N_\text{sub}\times N_\text{symb}}$, the channel 
\begin{equation}
    \Hm^{k,n} = \frac{\Ym^{k,n}}{\Xm^{k,n}}
\end{equation}
is estimated for each sub-element $k,n$, using the single-tap-equalizer. Exploiting that only a limited amount of paths $L$ is seen by the received, the channel can be rewritten into
\begin{equation}
\Hm^{k,n} = \sum_{\ell=0}^{L} p_{\text{loss}} \underbrace{e^{j2\pi nT_0f_{\ell}}}_{\text{Doppler}} +  \underbrace{e^{j2\pi k d_{\ell}/c_0 \Delta f}}_{\text{distance}}  + \mathcal{N}^{k,n},
\end{equation}
where the Doppler frequency shift $f_{\ell}$ of each path is creating a phase shift over the \ac{OFDM} symbols, with $T_0$ being the \ac{OFDM} symbol duration. The distance traveled creates a linear phase shift over the subcarriers with $\Delta f$ being the sub-carrier spacing. The zero-mean Gaussian distributed noise sample is given by $\mathcal{N}^{k,n}$. Thus, one can directly conclude that the phase-information per object can be used to determine the relative speed and the range of the object. The angle of the object is estimated by the phase-difference between antennas, i.e. having multiple $\Hm^{k,n}$. 
The baseline to exploit this orthogonality is represented by calculating the periodogram of the channel \cite{Braun2019}
\begin{equation}
\Pm^{k,n} = \left| \sum_{m=0}^{N_\text{smyb}-1}\left(\sum_{p=0}^{N_\text{sub}-1}\Hm^{p,m}e^{-j2\pi\frac{pn}{N_\text{symb}}}\right)e^{j2\pi\frac{mk}{N_\text{sub}}}\right|^2
\end{equation}
using the \ac{FFT} over the symbols and the inverse \ac{FFT} over the number of subcarriers. Thus, the phase-shifts create a peak at the corresponding distance and speed of the respective object.  This commonly used technique  has limitations regarding resolution and accuracy\cite{Braun2019}. 
Therefore, subspace methods were proposed to exploit the underlying channel covariance matrix 
\begin{equation}
    \Rm = \vec{\Hm}\vec{\Hm}^{\text{H}}.
\end{equation}
One prominent candidate is \ac{MUSIC} \cite{Poro2010}, as its exploits the noise subspace of the covariance matrix, which is calculated using the \ac{EVD} of $\Rm$ and partitioning the eigenvectors into the signal subspace $\mathbf{U}_S$ corresponding to the $\mathit{Q}$ strongest eigenvalues and the complementary noise subspace $\mathbf{U}_N$. This noise subspace is probed using the the corresponding steering vectors
\begin{align}
\mathbf{a}(\phi_q) &= \begin{bmatrix}
		1, \ e^{j2\pi  \frac{d}{\lambda} \sin(\phi_{q})},\ \dots, \ e^{j2\pi (N_\text{ant} -1) \frac{d}{\lambda} \sin(\phi_{q}) }
\end{bmatrix}^\text{T}  \\
\mathbf{b}(d_{q}) &= \begin{bmatrix}
       1, \ e^{-j2\pi \Delta f \cdot \frac{d_{q}}{c}}, \ \dots, \ e^{-j2\pi (N_\text{sub} - 1) \Delta f \cdot \frac{d_{q}}{c}}
\end{bmatrix}^\text{T} \\
\label{eq:channel_vectors}	
\end{align}
where $\av$ is the angular and $\bv$ the range steering vector, with  $f_{D}$ being the Doppler frequency-shift, $N_\text{ant}$ being the number of antennas of a uniform linear array with $\lambda/2$ spacing. The 2D \ac{MUSIC} spectrum (range-azimuth) is obtained by computing 
\begin{equation}
P_{\text{MU}}(d, \phi) = \frac{1}{(\mathbf{a}(\phi) \otimes \mathbf{b}(d))^\text{H} \textbf{U}_{N}\textbf{U}_{N}^\text{H}(\mathbf{a}(\phi) \otimes\mathbf{b}(d))},
\label{eq:MUSIC}
\end{equation}
where $\otimes$ is the Kronecker product. For now we consider the \ac{RADAR} plot always to be the 2D-\ac{MUSIC} equivalent.

\subsection{Challenges of sensing}
In this paper we assume that the co-located transmitter and receiver are synchronized, giving an upper bound for performance. 
\begin{figure}[t]
\subfloat[]{
      \includegraphics{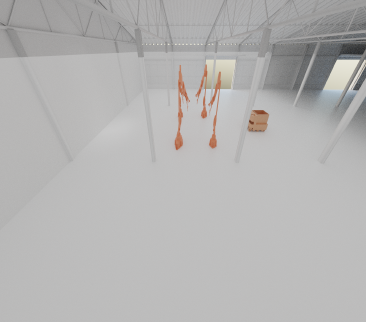}
  \label{fig:camera-scenario}
  }
   \subfloat[]{
       \includegraphics{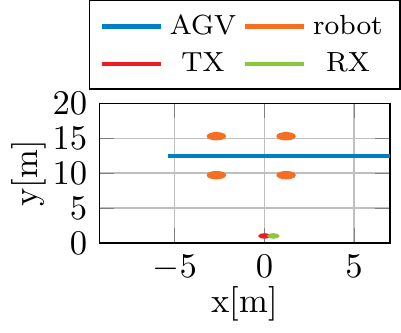}
   \label{fig:scenario}
  }
  \caption{Different outputs of MaxRay: \protect\subref{fig:camera-scenario} the rendered camera image snapshot
 \protect\subref{fig:scenario} he 2D experiment setup, with the \ac{AGV} moving with constant y coordinate.}
  \label{fig:ExampleScenario}    \vspace{-0.6cm}
\end{figure}
To show the algorithmic challenges we consider a static scenario first, e.g. Fig.~\ref{fig:camera-scenario} depicts for a specific time instant (frame) the camera and  Fig.~\ref{fig:scenario} shows the \ac{AGV} movement within the experiment. Note, the floor-map corresponds to a typical future production cell \cite{Arnold2021}, consisting of four robots in a specific configuration and an \ac{AGV} transporting materials within. 
For this investigation we use a carrier frequency of $f_c=\SI{3.75}{\giga \hertz}$, bandwidth of \SI{100}{\mega \hertz}, \ac{MIMO} configuration of $1\times4$ using a linear patch array, 100 symbols and 1024 subcarriers.
\begin{figure*}
  \subfloat[]{
      \includegraphics{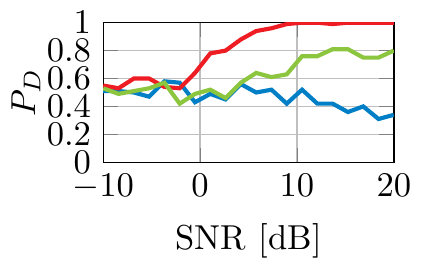}
  \label{fig:prob_static}
  }
   \subfloat[]{
       \includegraphics{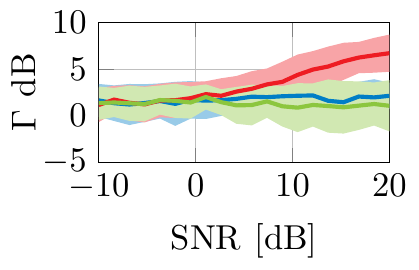}
   \label{fig:sinr_static}
  }
  \subfloat[]{
       \includegraphics{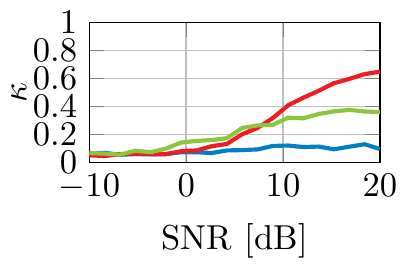}
    \label{fig:prom_static}
   }
     \subfloat[]{
         \includegraphics{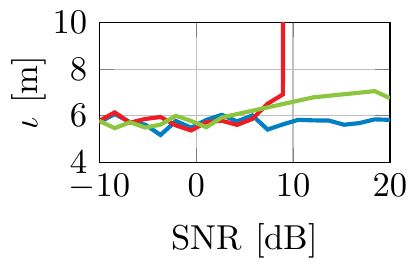}
    \label{fig:iso_static}
   }
  \caption{Comparison of different metrics for performance estimation of clutter removal. \protect\subref{fig:prob_static} Probability of detection.\protect\subref{fig:sinr_static} SINR \protect\subref{fig:prom_static} Prominence. \protect\subref{fig:iso_static} Isolation.}
  \label{fig:MetricsComparison}
  \vspace{-0.6cm}
\end{figure*}
\subsection{Clutter Removal}
Clutter is defined in general as the interference, noise and reflection from unwanted targets. The task of clutter rmoval is therefore to remove any signal not being impacted by the \ac{AGV}.
\subsubsection{Reference method}
This reference clutter removal method is based on measuring the average environment with and without the \ac{AGV} within $\Hm$ and $\Hm_{\text{ref}}$ respectively. Then, we subtract the reference $\Hm_{\text{ref}}$ from the measurement.
%\begin{equation}
%\hat{\Hm} = \text{FFT}_{N_\text{sub}}\left[\text{IFFT}_{N_\text{sub}}\left[\Hm\right] - \text{IFFT}_{N_\text{sub}}\left[\Hm_\text{ref}\right] \right].
%\end{equation}

Fig.~\ref{fig:exmaple-diff-clutter} depicts left the original 2D-\ac{MUSIC} plot and in the middle the output of clutter removal using the reference method. Thus, the wanted target (\ac{AGV}) at a range of $\approx$ \SI{28}{\metre} and 50\textdegree $ $ is clearly visible. It can be shown that the clutter removal works very good if the scenario is kept.

\subsubsection{Dynamic method}
\begin{figure}[t]
    \centering
        \includegraphics{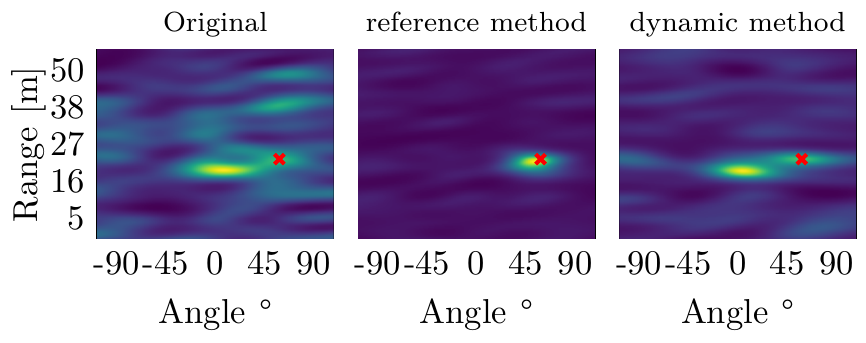}
    \caption{Example image of clutter removal, left original, middle clutter removed using the reference method and right the dynamic method}
    \label{fig:exmaple-diff-clutter}
     \vspace{-0.6cm}
\end{figure}

Another method to remove clutter and to detect a moving object is to estimate the phase shift over time of the individual impulses, by
\begin{equation}
\Delta \hv  = \sum_{p=0}^{N_\text{sub}-1}\Hm^{p,0}e^{\frac{-j2\pi p}{N_\text{sub}}}- \sum_{p=0}^{N_\text{sub}-1}\Hm^{p,N_\text{symb}}e^{\frac{-j2\pi p}{N_\text{sub}}}.
\end{equation}
Using the condition that an impulse not affect by movement should have no phase difference over the frame (besides noise), the impulses in the time domain are set to zero with $\Delta \hv \leq \epsilon$, where $\epsilon$ is arbitrarily set to 1\%. 
Fig.~\ref{fig:exmaple-diff-clutter} shows the impact of this clutter removal technique (right), removing partly the clutter. Thus, this technique enhances the \ac{RADAR} image, by de-cluttering.
The importance of finding a suitable metric to compare those two clutter-removal techniques is emphasized by understanding that the visual effect does not give any possibility of numerically assess their performance, for fine-granular comparisons.

\subsection{Metrics}
Although in vision-based systems different metrics were introduced, \ac{RADAR} images have the unique twist of not knowing if and how 
\begin{itemize}
    \item the target is "seen" by the receiver (there is a reflection),
    \item the size of the objects impacts  the receiver image,
    \item the exact position and reflection are accounted for.
\end{itemize}
To demonstrate the advantages and disadvantages of this techniques we leverage first a static frame, where the ground-truth is known. In general, we assume that peaks can be detected if their amplitude/power is at least 10\% higher power than the average power of the \ac{RADAR} image. We consider now four different metrics, where the probability of detection emerged from vision technologies, the \ac{SINR} metric from communication, the prominence and isolation from geology. 

\subsubsection{Probability of Detection}
The probability of detection is defined as
\begin{equation}
   P_D = \frac{N_\text{detected}}{N_\text{iteration}} \cdot 100 \%
\end{equation}
where $N_\text{detected}$ is the number how often the \ac{AGV} was detected within a range of $\lambda$ around the true target and $N_\text{iteration}$ is the number of the experiment runs. Plotting this metric over the movement of the \ac{AGV} demonstrates potential blockers and thus can be leveraged to enhance communication. This metric is also heavily impacted by defining the range difference allowed, making erroneous peaks possible in low \ac{SNR} regions. 

\subsubsection{SINR}
Another metric is the \ac{SINR}
\begin{equation}
    \Gamma = \frac{P_\text{Signal}}{\sum P_{S}},
\end{equation}
where the power at the specific position in the \ac{RADAR} image $P_\text{Signal}$ is taken and then divided by the remaining power $P_S$ in the set $S$ outside the $\lambda$ circle. Note, this metric incorporates targets which are not wanted as interference and thus punishes  clutter-removal techniques which keep background targets (dynamic method).

\subsubsection{Prominence}
Prominence
\begin{equation}
    \kappa = \frac{P_\text{Signal}}{P_\text{c}}
\end{equation}
 is a metric mostly used in topography maps and is defined as the distance between the wanted signal (peak) $P_\text{Signal}$ and the circumference $P_\text{c}$ around this peak, where the gradient changes its sign. 
\subsubsection{Isolation}
Isolation
\begin{equation}
\iota = \left|\pv_\text{peak} - \pv_\text{closest,peak} \right|^2
\end{equation}
is given as the distance between the signal peak $\pv_\text{peak}$ to the next peak $\pv_\text{closest,peak}$. Note that this metric is unbounded if only one target is available.
Fig.~\ref{fig:MetricsComparison} depicts in this simple case the performance  for the different clutter removal techniques using the mentioned metrics. It can be seen that since the noise can create erroneous peaks at the wanted position, the probability of detection is falsely large at low \ac{SNR}, but converges to the real probability in high \ac{SNR} regions. As seen from the visualization in the clutter removal section the reference case is better than the dynamic removal but still achieves a lot better results than without. Thus, both techniques seem to be viable at first. The \ac{SINR} metric punishes that the second peak of the clutter is not removed, thus no real gain is shown and it seems to be an unsuitable metric (comp. Fig.~\ref{fig:exmaple-diff-clutter}). The prominence in the sub-figure does not show the effect of the low \ac{SNR} error as the prominence in noise is almost zero. Further it shows the respective gains of the clutter removal, rendering it into a very viable metric for sensing algorithms.  The isolation in the last sub-figure shows the gains, but is unbounded due to only a remaining peak, when the SINR is high enough.
In conclusion it seems that prominence due to its bounding by zero and one, incorporates all underlying effects and is suitable for the most classical and \ac{DL} techniques.

%% file: sections/simulation.tex
After knowing the behaviour of the probability of detection and prominence in the static case, we move further to another experiment where we run the \ac{AGV} on the track shown in the Fig.~\ref{fig:scenario} and investigate the time behaviour (blockage)
 and the changes if one of the robots moves (environmental movement).
\subsection{Dynamic environment}
\begin{figure}[t]
    \centering
       \includegraphics{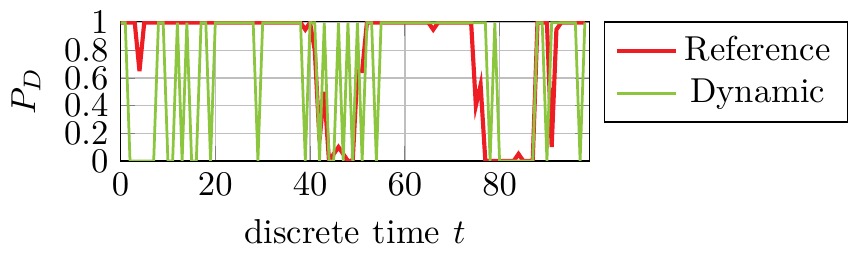}
    \caption{Probability of detection over time for the two clutter removal techniques}
        \vspace{-0.2cm}
    \label{fig:prob_over_time}
    \vspace{-0.4cm}
\end{figure}
Fig.~\ref{fig:prob_over_time} shows the probability of detecting the \ac{AGV} using the two clutter removal techniques. It can be seen that the reference case only has two dips at exactly the positions of the robotic arm (e.g. blockage). Further due to the limited angular resolution the \ac{AGV} cannot be resolved.
In the case of the dynamic clutter removal the channel impulse response in some cases cannot resolve the target, thus making it unsuitable for perfect tracking. Note that time tracking can enhance this method. 
In this case communication can benefit from sensing systems by predicting blockage, and therefore performance drops, beforehand.

\subsection{Environmental movement}
\begin{figure}[t]
    \centering
        \includegraphics{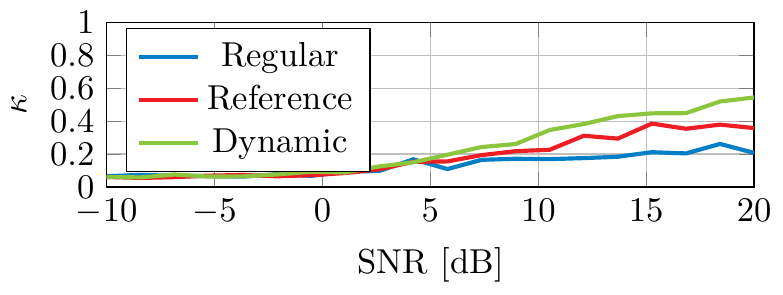}
    \caption{Average performance of clutter removal algorithms in the case of environmental movement.}
       \vspace{-0.2cm}
    \label{fig:environmental_change}     \vspace{-0.4cm}
\end{figure}

Fig.~\ref{fig:environmental_change} depicts the performance, if one of the robots arm moves, while the \ac{AGV} moves, degrading the performance of the reference method. This is due to the effect that the reference method only incorporates a specific setting and the reflections within the environment drastically change if the environment changes. Notably the performance of the dynamic clutter removal stays constant and even improves a little, due to better separation of static parts. Further the dynamic case seems to be achieving the most promising results, which could be used to calculate the \ac{RADAR} image.

%% file: sections/conclusion.tex
\section{Conclusion}
A versatile framework capable of simulating \ac{ISAC} systems using realistic scenarios due to the virtue of Blender was introduced. The required workflow of such  a sensing framework to incorporate all necessary propagation attributes was demonstrated.
Further four commonly used metrics from different fields were compared  regarding their ability to capture the sensing capabilities. As it turns out prominence due to its noise resilience and being bounded is a strong candidate for creating a common comparison technique among a large amount of different algorithms.
In a standard industrial use-case state-of-the art clutter removal techniques were compared, demonstrating that clutter removal is a key element for the success of \ac{ISAC} systems. 
Additional scenarios and datasets will be provided in the future to reproduce and compare algorithms.

%% file: main.bbl
\begin{thebibliography}{10}
\providecommand{\url}[1]{#1}
\csname url@samestyle\endcsname
\providecommand{\newblock}{\relax}
\providecommand{\bibinfo}[2]{#2}
\providecommand{\BIBentrySTDinterwordspacing}{\spaceskip=0pt\relax}
\providecommand{\BIBentryALTinterwordstretchfactor}{4}
\providecommand{\BIBentryALTinterwordspacing}{\spaceskip=\fontdimen2\font plus
\BIBentryALTinterwordstretchfactor\fontdimen3\font minus
  \fontdimen4\font\relax}
\providecommand{\BIBforeignlanguage}[2]{{%
\expandafter\ifx\csname l@#1\endcsname\relax
\typeout{** WARNING: IEEEtran.bst: No hyphenation pattern has been}%
\typeout{** loaded for the language `#1'. Using the pattern for}%
\typeout{** the default language instead.}%
\else
\language=\csname l@#1\endcsname
\fi
#2}}
\providecommand{\BIBdecl}{\relax}
\BIBdecl

\bibitem{TWild2021}
T.~Wild, V.~Braun, and H.~Viswanathan, ``{Joint Design of Communication and
  Sensing for Beyond 5G and 6G Systems},'' \emph{IEEE Access}, vol.~9, pp.
  30\,845--30\,857, 2021.

\bibitem{Lima2021}
{C. De Lima, et. al.}, ``{Convergent Communication, Sensing and Localization in
  6G Systems: An Overview of Technologies, Opportunities and Challenges},''
  \emph{IEEE Access}, vol.~9, pp. 26\,902--26\,925, 2021.

\bibitem{viswanathan2020communications}
H.~Viswanathan and P.~E. Mogensen, ``Communications in the 6g era,'' \emph{IEEE
  Access}, vol.~8, pp. 57\,063--57\,074, 2020.

\bibitem{Braun2019}
M.~Braun, C.~Sturm, and F.~Jondral, ``{On the Frame Design for Joint OFDM Radar
  and Communications},'' \emph{Vehicular Technology Conference (VTC) Spring
  2010}, 01 2010.

\bibitem{Leen2012}
L.~Sit, C.~Sturm, J.~Baier, and T.~Zwick, ``{Direction of arrival estimation
  using the MUSIC algorithm for a MIMO OFDM Radar},'' \emph{2012 IEEE Radar
  Conference}, pp. 1--7, 2012.

\bibitem{saur20205gcar}
{Stephan Saur, et. al.}, ``5gcar demonstration: Vulnerable road user protection
  through positioning with synchronized antenna signal processing,'' in
  \emph{WSA 2020; 24th International ITG Workshop on Smart Antennas}.\hskip 1em
  plus 0.5em minus 0.4em\relax VDE, 2020, pp. 1--5.

\bibitem{Jian2020}
J.~Liu, H.~Liu, Y.~Chen, Y.~Wang, and C.~Wang, ``{Wireless Sensing for Human
  Activity: A Survey},'' \emph{{IEEE Communications Surveys Tutorials}},
  vol.~22, no.~3, pp. 1629--1645, 2020.

\bibitem{Wenfeng2015}
W.~He, K.~Wu, Y.~Zou, and Z.~Ming, ``{WiG: WiFi-Based Gesture Recognition
  System},'' in \emph{{2015 24th International Conference on Computer
  Communication and Networks (ICCCN)}}, 2015, pp. 1--7.

\bibitem{Arnold2019}
M.~Arnold, J.~Hoydis, and S.~t. Brink, ``{Novel Massive MIMO Channel Sounding
  Data applied to Deep Learning-based Indoor Positioning},'' in \emph{{SCC
  2019; 12th International ITG Conference on Systems, Communications and
  Coding}}, 2019, pp. 1--6.

\bibitem{3GPP}
3GPP, ``{Study on channel model for frequencies from 0.5 to 100 GHz},'' {3rd
  Generation Partnership Project (3GPP)}, Technical Report (TR) 38.901, Nov.
  2021, version 16.1.0.

\bibitem{Blender}
\BIBentryALTinterwordspacing
{Blender Online Community}, \emph{Blender - a 3D modelling and rendering
  package}, Blender Foundation, 2018. [Online]. Available:
  \url{http://www.blender.org}
\BIBentrySTDinterwordspacing

\bibitem{Rappaport2019}
T.~S. Rappaport, Y.~Xing, O.~Kanhere, S.~Ju, A.~Madanayake, S.~Mandal,
  A.~Alkhateeb, and G.~C. Trichopoulos, ``{Wireless Communications and
  Applications Above 100 GHz: Opportunities and Challenges for 6G and
  Beyond},'' \emph{IEEE Access}, vol.~7, pp. 78\,729--78\,757, 2019.

\bibitem{ITU}
ITU, ``{Propagation by diffraction - Standards},'' { International
  Telecommunication Union (ITU)}, Technical Specification (TS) ITU-R P.526-1,
  01 2018.

\bibitem{Poro2010}
M.~G. Porozantzidou and M.~T. Chryssomallis, ``{Azimuth and elevation angles
  estimation using 2-D MUSIC algorithm with an L-shape antenna},'' in
  \emph{2010 IEEE Antennas and Propagation Society International Symposium},
  2010, pp. 1--4.

\bibitem{Arnold2021}
{M.~Arnold, P.~Baracca, T.~Wild, F.~Schaich, S.~ten Brink}, ``{Measured
  Distributed vs Co-located Massive MIMO in Industry 4.0 Environments},'' in
  \emph{2021 Joint European Conference on Networks and Communications 6G Summit
  (EuCNC/6G Summit)}, 2021, pp. 306--310.

\end{thebibliography}
